\documentclass{webofc}
\usepackage[varg]{txfonts}   % Web of Conferences font

\usepackage{lineno}

\begin{document}
\title{Overview of open heavy-flavour and quarkonia measurements with ALICE}
%
% subtitle is optionnal
%
%%%\subtitle{Do you have a subtitle?\\ If so, write it here}

\author{\firstname{Andrea Dubla}\inst{1}\fnsep\thanks{\email{andrea.dubla@cern.ch}} \lastname{for the ALICE Collaboration} }

\institute{GSI Helmholtzzentrum f{\"u}r Schwerionenforschung, Planckstra{\ss}e 1, 64291 Darmstadt, Germany 
          }

\abstract{%

Heavy-flavour hadrons, i.e. hadrons containing charm or beauty quarks, are effective probes to test perturbative-QCD (pQCD) calculations, to investigate the different hadronisation mechanisms, and to study the quark--gluon plasma (QGP) produced in relativistic heavy-ion collisions at the LHC. 
Measurements performed in pp and p--Pb collisions have recently revealed unexpected features not in line with the expectations
based on previous measurements from $\rm{e^+e^-}$ and ep collisions, showing that charm fragmentation fractions are not universal. The investigation of initial-state effects such as shadowing in the collision of a proton with a heavy nucleus is also performed. Measurements of open heavy-flavour and quarkonia production in Pb--Pb collisions allow for testing the mechanisms of heavy-quark transport, energy loss, and coalescence effects during the hadronisation in the presence of a QCD medium. In this contribution, the most recent results on open heavy-flavour and quarkonia production in pp, p--Pb, and Pb--Pb collisions obtained by the ALICE Collaboration are discussed.}
\maketitle
\section{Introduction}
\label{intro}
The study of heavy-flavour hadron production in proton--proton (pp) collisions provides an important test for quantum chromodynamics (QCD) calculations. Heavy-flavour hadron production is usually computed with pQCD calculations as the convolution of the parton distribution functions (PDFs) of the incoming protons, the partonic cross section and the fragmentation functions that describe the transition from charm quarks into charm hadrons. The latter are typically parametrised from measurements performed in  $\rm e^+e^-$ or ep collisions, under the assumption that the hadronisation of charm quarks into charm hadrons is a universal process independent of the colliding system.
In the presence of a quark--gluon plasma (QGP), the strongly-interacting colour-deconfined state of matter created in ultra-relativistic heavy-ion collisions, an additional hadronisation mechanism, alternative to the in-vacuum string fragmentation, is considered. In this mechanism, known as coalescence, light quarks from the medium coalesce with the heavy quark to form a meson or baryon. By studying heavy-flavour production in nucleus--nucleus collisions, the relevance of coalescence in the medium for heavy quarks can be probed. Models that include hadronisation via coalescence for charm quarks qualitatively describe D-meson measurements in Pb--Pb collisions at the LHC. The possible occurrence of coalescence in pp collisions is one of the topics of investigation to provide a better description of heavy-baryon production.

\section{Charm production and fragmentation in pp collisions}
\label{sec-1}
\begin{figure*}[ht!]
\centering
\includegraphics[width=.29\textwidth]{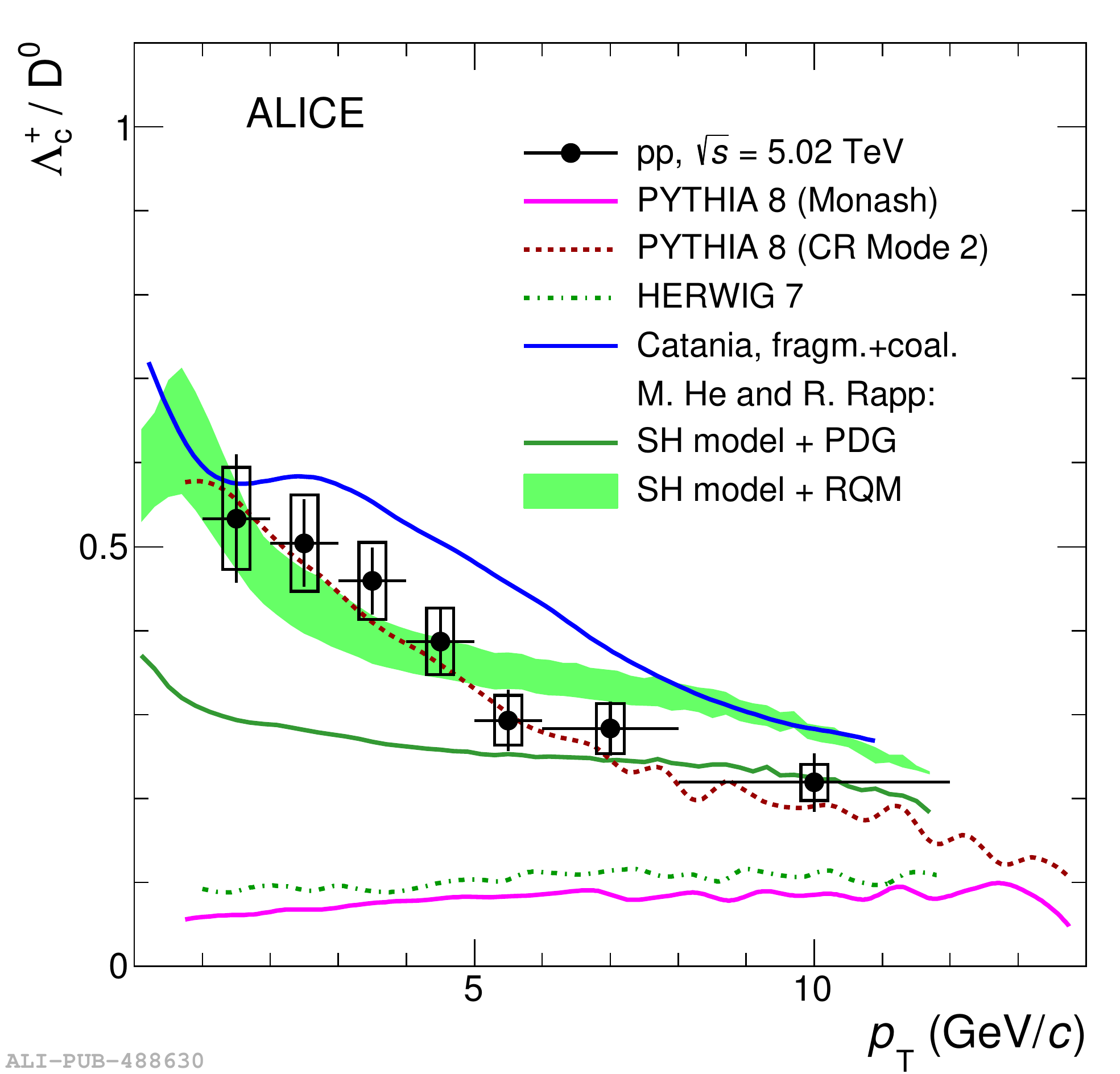}
\includegraphics[width=.32\textwidth]{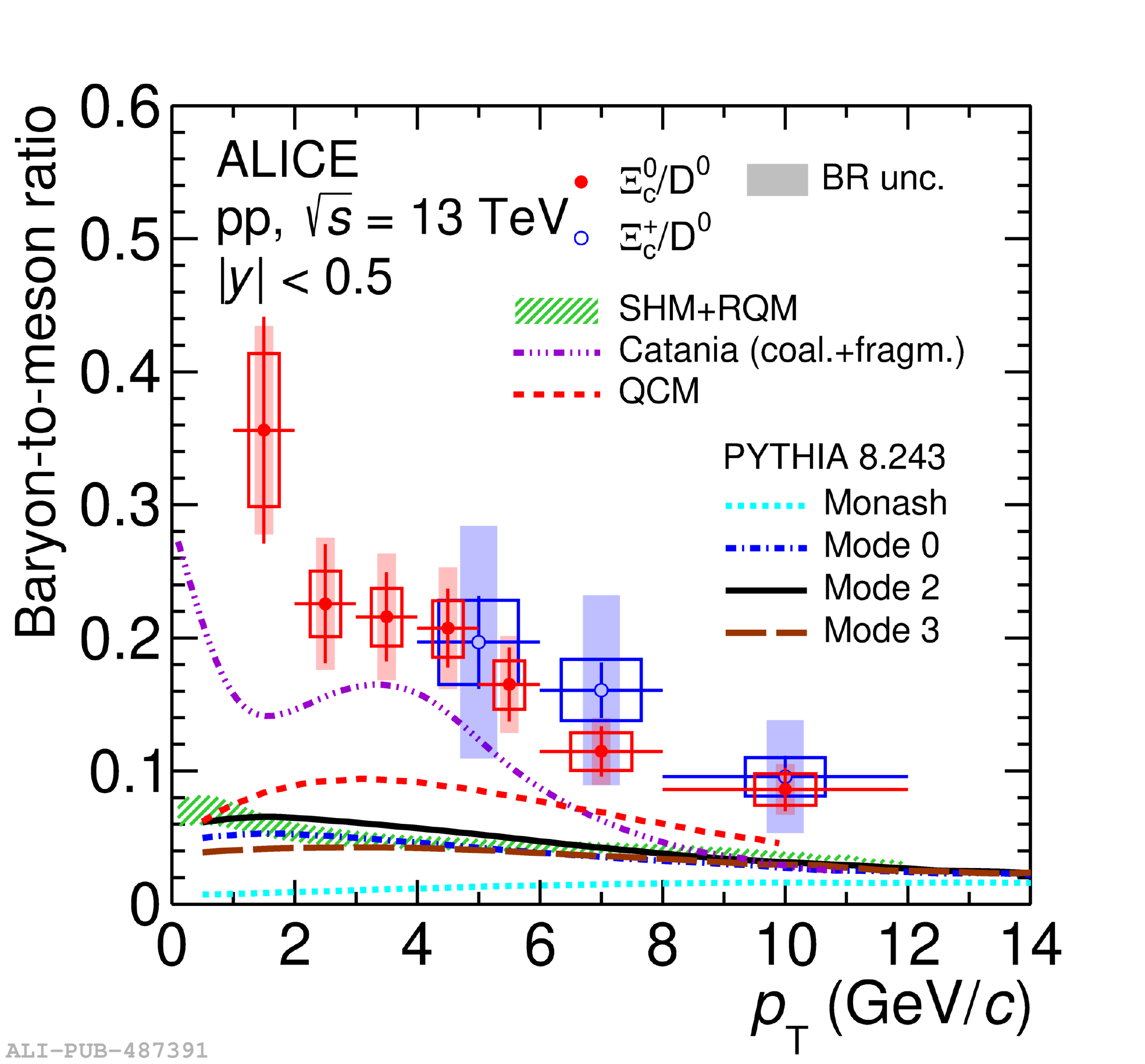}
\includegraphics[width=.33\textwidth]{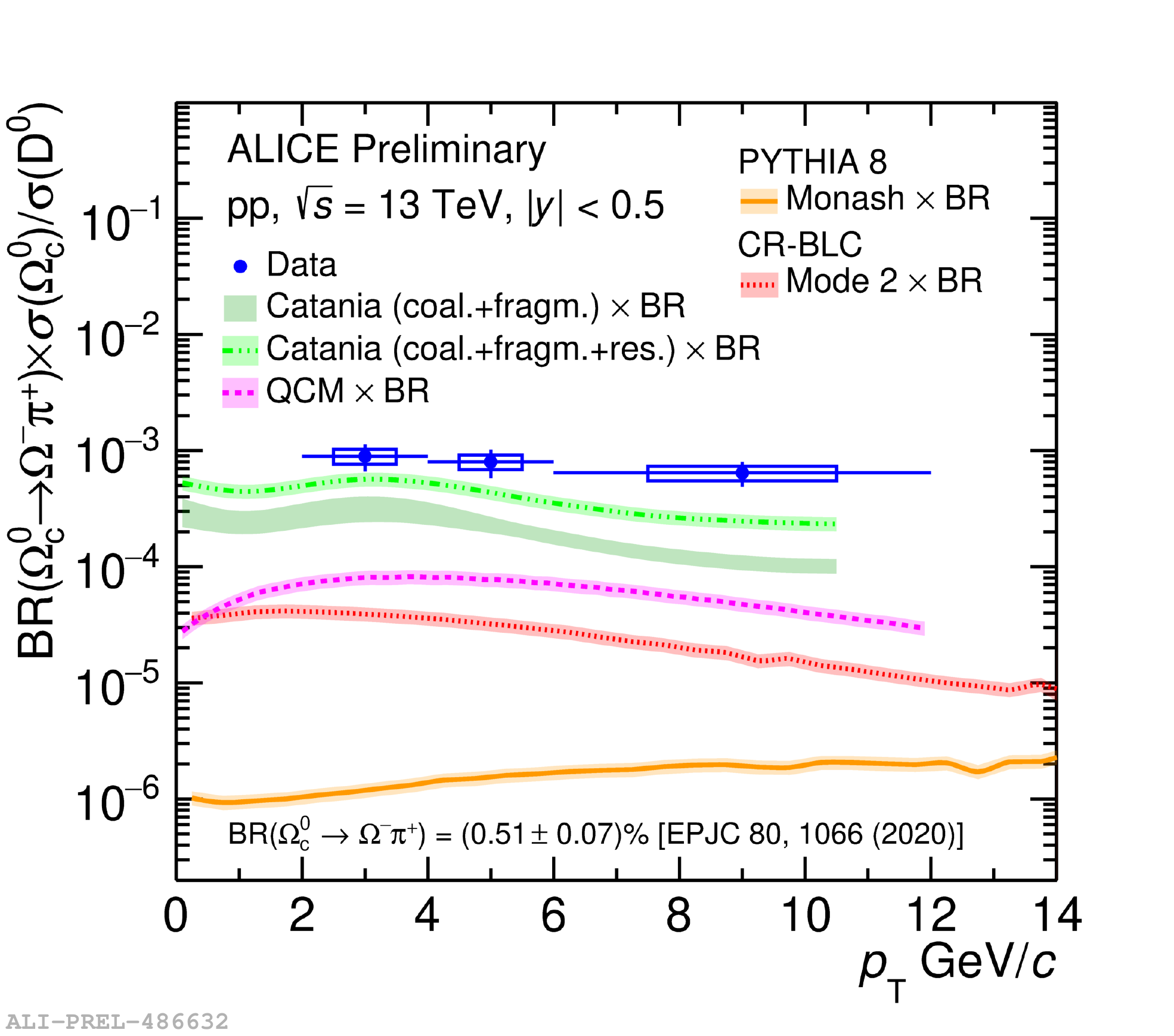}
\caption{Left panel: The $\Lambda_{\rm c}^+/{\rm D^0}$ ratio measured in pp collisions at  $\sqrt{s}$ = 5.02 TeV~\cite{Acharya:2020uqi}. Middle panel: The $\rm \Xi_c^0/D^0$ and $\rm \Xi_c^+/D^0$ ratios as a function of $p_{\rm T}$ in pp collisions at $\sqrt{s}=13$~TeV~\cite{xic13tev}. Right panel: The ${\rm BR}\times\Omega^0_{\rm c}/{\rm D^0}$ as a function of $p_{\rm T}$ in pp collisions at $\sqrt{s}=13$~TeV. All yield ratios are compared to theoretical model calculations.}
\label{fig-2}       % Give a unique label
\end{figure*}

The large data samples collected during Run 2 of the LHC at $\sqrt{s}$ = 5.02 and 13 TeV allowed ALICE to measure the vast majority of charm quarks produced at midrapidity in the proton-proton (pp) collisions by reconstructing the decays of the ground-state charm hadrons down to very low transverse momenta ($p_{\rm T}$), i.e. all charm-meson species (${\rm D^0}$, ${\rm D^+}$, ${\rm D_s^+}$~\cite{Acharya:2021cqv}) and charm-baryon species ($\Lambda_{\rm c}^+$~\cite{Acharya:2020uqi}, $\Xi_{\rm c}^{0,+}$~\cite{Acharya:2021dsq,xic13tev} and $\Omega^0_{\rm c}$).
The ratios of different hadron species as function of the transverse momentum $p_{\rm T}$ help to study heavy flavour hadronisation. Both the prompt and non-prompt (coming from B-hadron decays) meson-to-meson yield ratios of D mesons are compatible with pQCD calculations~\cite{Cacciari:2012ny,Kramer:2017gct,Helenius:2018uul} using fragmentation fractions extracted from $\rm e^+e^-$ collision data. Such calculations, however, significantly underpredict all the measured baryon-to-meson yield ratios, indicating that the fragmentation fractions are not universal. The $\Lambda_{\rm c}^+/{\rm D^0}$ ratio in pp collisions, reported in the left panel of Fig.~\ref{fig-2},  shows a clear decrease with increasing $p_{\rm T}$ and it is better described by models with an extension of colour reconnection beyond the leading colour approximation~\cite{Christiansen:2015yqa}, models relying on hadronisation
via coalescence~\cite{Song:2018tpv,Minissale:2020bif}, or statistical hadronisation models with an augmented set of baryon states predicted by the relativistic quark
model (RQM)~\cite{He:2019tik}.
Measurements of the heavier charm-strange baryons $\Xi_{\rm c}^{0,+}$ and $\Omega^0_{\rm c}$  pose further important constraints to charm-quark hadronisation models. For the $\Omega^0_{\rm c}$ the absolute branching ratio (BR) of the decay channel $\Omega^0_{\rm c} \rightarrow \Omega^\pm \pi^\mp$ is not measured, hence only the BR multiplied with the cross section is reported. The yield ratios of these charm baryons to the ${\rm D^0}$-meson one are presented in the middle and right panel of Fig.~\ref{fig-2}, and compared to predictions from the same models that describe the $\Lambda_{\rm c}^+/{\rm D^0}$ ratio.
For the $\Xi_{\rm c}^{0,+}/{\rm D^0}$ and ${\rm BR}\times\Omega^0_{\rm c}/{\rm D^0}$ all calculations, except the Catania model~\cite{Minissale:2020bif}, which gets closer to the data and relies on hadronisation via coalescence, are underpredicting the measured ratios, hinting to an even stronger enhancement for charm-strange baryons.
The predictions tuned on measurements in $\rm e^+e^-$ collisions underestimate all three baryon-to-meson yield ratios~\cite{Skands:2014pea}, providing further evidence that different processes are involved in charm hadronisation for elementary and hadronic collisions.
These new charm-baryon measurements provide important constraints to models of charm-quark hadronisation in pp collisions, being, in particular, sensitive to the description of charm-strange baryon production in the colour reconnection approach, and to the possible contribution of coalescence to charm-quark hadronisation in pp collisions.

\begin{figure*}[ht!]
\centering
\includegraphics[width=.4\textwidth]{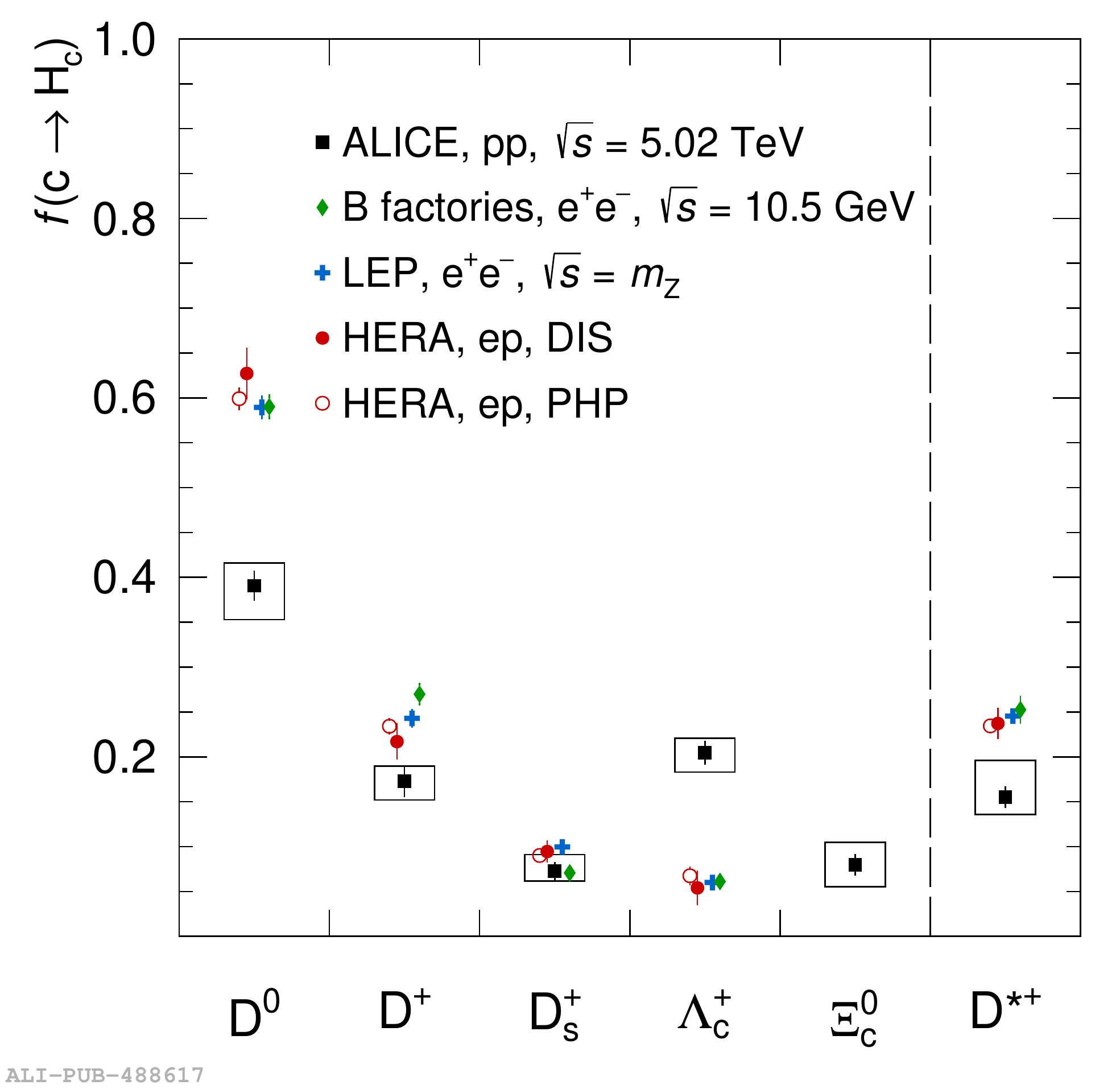}
\includegraphics[width=.4\textwidth]{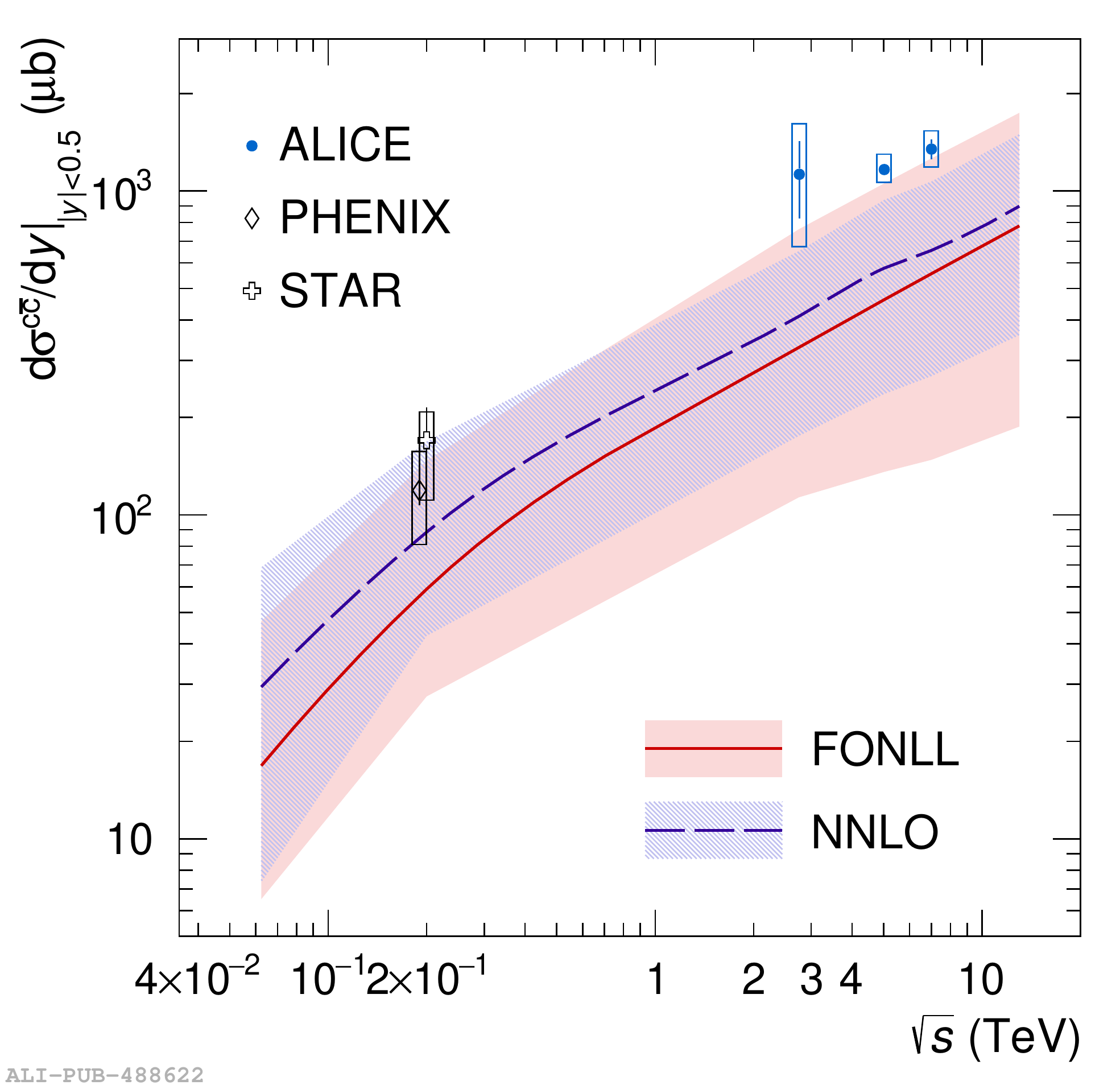}
\caption{Left panel: Charm-quark fragmentation fractions into charm hadrons measured in pp collisions at $\sqrt{s}$ = 5.02 TeV in comparison with experimental measurements performed in ${\rm e^+e^-}$ and in ep collisions~\cite{ALICE:2021dhb}. Right panel: Charm production cross section at midrapidity per unit of rapidity as a function of the collision energy. The FONLL and  NNLO calculations are also shown~\cite{Cacciari:2012ny,dEnterria:2016yhy,Czakon:2013goa}.}
\label{fig-3}       % Give a unique label
\end{figure*}

The charm fragmentation fractions, $f({\rm c} \rightarrow {\rm H_c})$ (shown in the left panel of Fig.~\ref{fig-3}), which represent the probabilities of a charm quark to hadronise into a given charm hadron, were measured for the first time including baryon species at midrapidity in pp collisions at the LHC and are observed to be different from the ones measured in $\rm e^+e^-$ and ep collisions, showing evidence that the assumption of universality (collision-system independence) of charm-to-hadron fragmentation is not valid~\cite{ALICE:2021dhb}. The charm quarks hadronize into baryons almost 40\% of the time, which is four times more often than what was measured at colliders with electron beams. The fragmentation fraction for the $\Xi_{\rm c}^0$ was measured for the first time in any collision system. 
The ${\rm c\bar{c}}$ production cross section per unit of rapidity at midrapidity ($\rm{d} \sigma^{c\bar{c}}/\rm{d}y|_{|y| < 0.5}$ ) was calculated by summing the cross sections of all measured ground-state charm hadrons (${\rm D^0}$, ${\rm D^+}$, ${\rm D_s^+}$, $\Lambda_{\rm c}^+$, and $\Xi_{\rm c}^0$ and their charge conjugates). The contribution of the $\Xi_{\rm c}^{0}$ was multiplied by a factor of two, in order to account for the contribution of the $\Xi_{\rm c}^+$. The resulting ${\rm c\bar{c}}$  cross section per unit of rapidity at midrapidity is ${\rm d}\sigma^{\rm c\bar{c}}/{\rm d}y|_{|y| < 0.5} = 1165 \pm 44(\rm{stat}) ^{+134}_{-101}(\rm{syst})~\mu b$. This measurement was done for the first time in hadronic collisions at the LHC including the charm baryon states~\cite{ALICE:2021dhb}.
The newly measured fragmentation fractions at midrapidity in pp collisions at $\sqrt{s}$ = 5.02 TeV allowed the recomputation of the charm production cross sections per unit of rapidity at midrapidity in pp collisions at $\sqrt{s}$ = 2.76 and 7 TeV. The charm cross sections were obtained by scaling the $p_{\rm T}$-integrated ${\rm D^0}$-meson cross section for the relative fragmentation fraction of a charm quark into a ${\rm D^0}$ meson. The updated ${\rm c\bar{c}}$ cross sections at $\sqrt{s}$ = 2.76 and 7 TeV are about 40\% higher than the previously published results, reflecting the differences in the fragmentation into charm baryons with respect to those measured in ${\rm e^+e^-}$ and pp collisions. 
In  the  right  panel  of  Fig.~\ref{fig-3},  the  measured  ${\rm c\bar{c}}$ production cross sections per unit of rapidity at midrapidity  are  compared  with  FONLL~\cite{Cacciari:2012ny}  and  NNLO~\cite{dEnterria:2016yhy,Czakon:2013goa} predictions as a function of the collision energy. The ${\rm c\bar{c}}$  cross sections measured at midrapidity at the LHC lies at the upper edge of the theoretical pQCD calculations. The measurements described above not only provide constraints to pQCD calculations but are also important as references for the investigation of the charm-quark interaction with the medium created in heavy-ion collisions.

\section{Radial flow, hadronisation, and shadowing in p--Pb collisions}
\label{sec-2}

\begin{figure*}[ht!]
\centering
 \includegraphics[width=.35\textwidth]{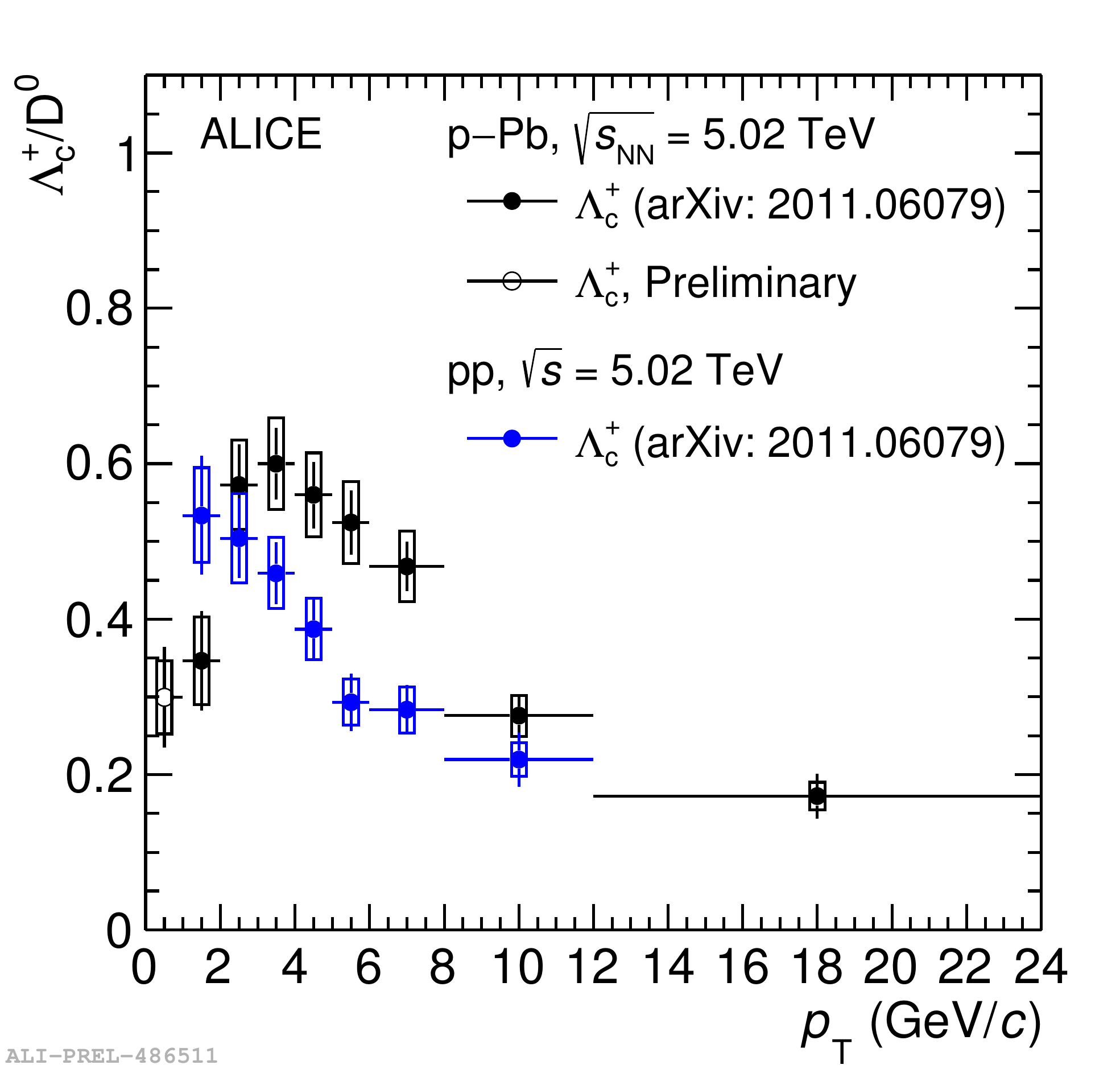}
\includegraphics[width=.58\textwidth]{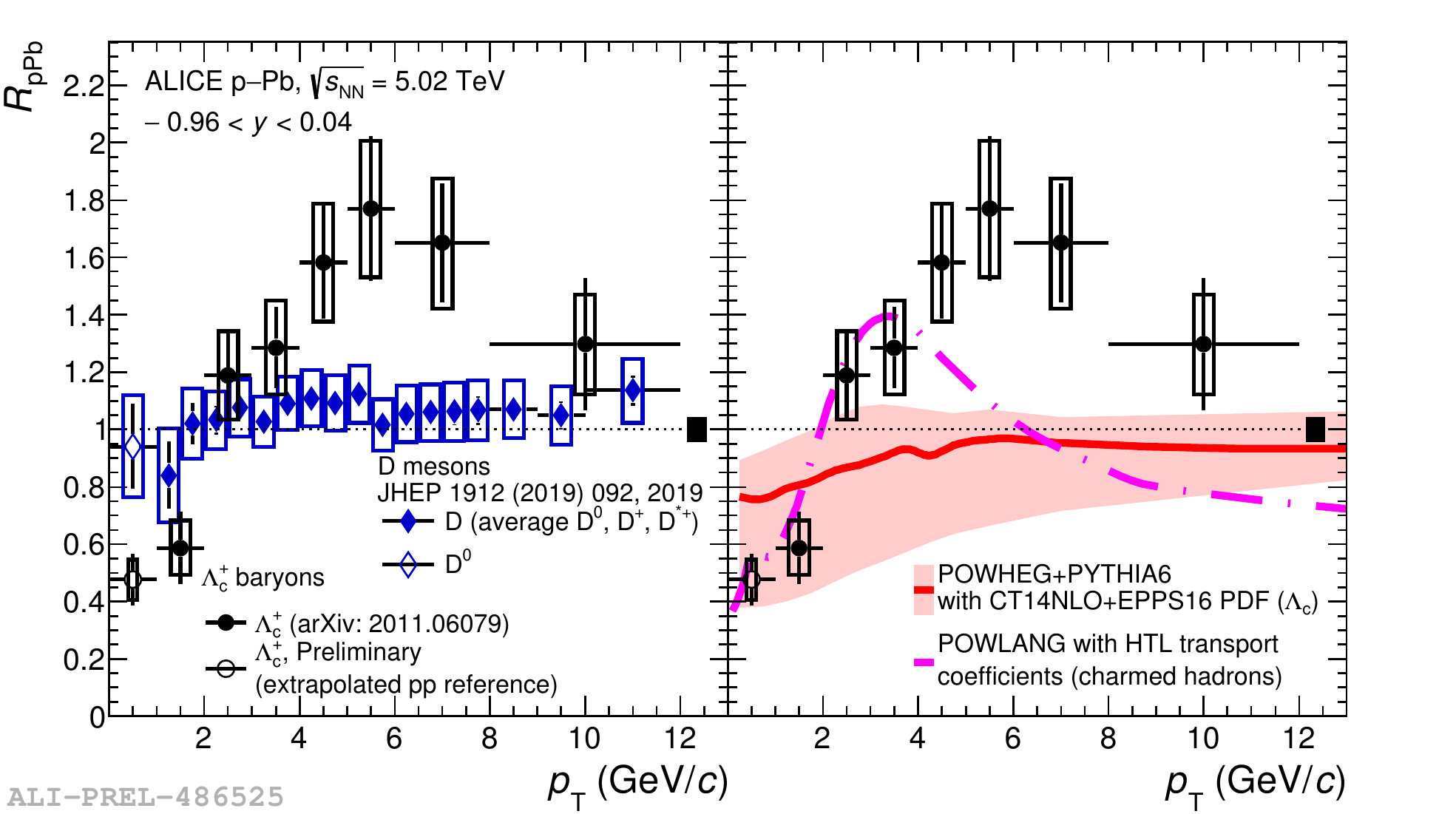}
\caption{Left panel: The $\Lambda_{\rm c}^+/{\rm D^0}$ yield ratio as a function of $p_{\rm T}$ measured in pp and in p--Pb collisions
 $\sqrt{s_{\rm NN}}$ = 5.02 TeV. Right panel: $R_{\rm pPb}$ of prompt $\Lambda_{\rm c}^+$ baryons in p--Pb collisions at  $\sqrt{s_{\rm NN}}$ = 5.02 TeV as a function of $p_{\rm T}$ compared to D-meson $R_{\rm pPb}$ (left) and to model expectations (right)~\cite{Acharya:2020uqi}.}%, compared to the $R_{\rm pPb}$ of D mesons as well as to POWHEG+PYTHIA 6~\cite{} with EPPS16~\cite{} simulations, and POWLANG~\cite{} predictions.}
\label{fig-4}       % Give a unique label
\end{figure*}

Additional investigations on the charm hadronisation measuring charm baryon production are carried out in p--Pb collisions at $\sqrt{s_{\rm NN}}$ = 5.02 TeV~\cite{Acharya:2020uqi}.
The $\Lambda_{\rm c}^+/{\rm D^0}$ yield ratio as a function of $p_{\rm T}$ in pp and p--Pb collisions is shown in the left panel of Fig.~\ref{fig-4}. For the first time at the LHC the $\Lambda_{\rm c}^+$ has been measured down to $p_{\rm T}$ = 0. A clear decreasing trend with increasing $p_{\rm T}$ is seen in p--Pb collisions for  $p_{\rm T} >$  2 GeV/$c$. The ratios measured in pp and p--Pb collisions are qualitatively consistent with each other, although a larger $\Lambda_{\rm c}^+/{\rm D^0}$ ratio in 3 $< p_{\rm T} <$ 8 GeV/$c$ and a lower ratio in 1 $< p_{\rm T} <$  2 GeV/$c$ are measured in p--Pb collisions with respect to pp collisions suggesting the presence of possible radial flow effect or of a further modification of the hadronisation mechanism in p--Pb collisions. 
In order to further investigate these effects and to study also cold nuclear matter effects (CNM), the nuclear modification factor ($R_{\rm pPb}$) was measured (right panel of Fig.~\ref{fig-4}). For $p_{\rm T} <$ 2 GeV/c an $R_{\rm pPb}$ lower than unity is measured  with a significance of 2.6$\sigma$, which provides a hint for a suppression of the yield in p--Pb with respect to pp collisions~\cite{Acharya:2020uqi}. The suppression is described by model calculations that include modification of the parton distribution function in Pb nuclei (shadowing). 
For $p_{\rm T} >$ 2 GeV/$c$ the $R_{\rm pPb}$ is systematically above unity, with a maximum deviation from $R_{\rm pPb}$ = 1 reaching 2.2$\sigma$ for 5 $< p_{\rm T} <$ 6 GeV/$c$. The indication that $\Lambda_{\rm c}^+$  production is suppressed at low $p_{\rm T}$ and enhanced at mid $p_{\rm T}$ in p--Pb collisions with respect to pp collisions could also be seen as possible radial flow effects present in p--Pb collisions. The current precision of the measurement is not sufficient to draw conclusions on the role of different CNM effects and the possible presence of final-state effects, like the radial flow.

\begin{figure*}[ht!]
\centering
\includegraphics[width=.45\textwidth]{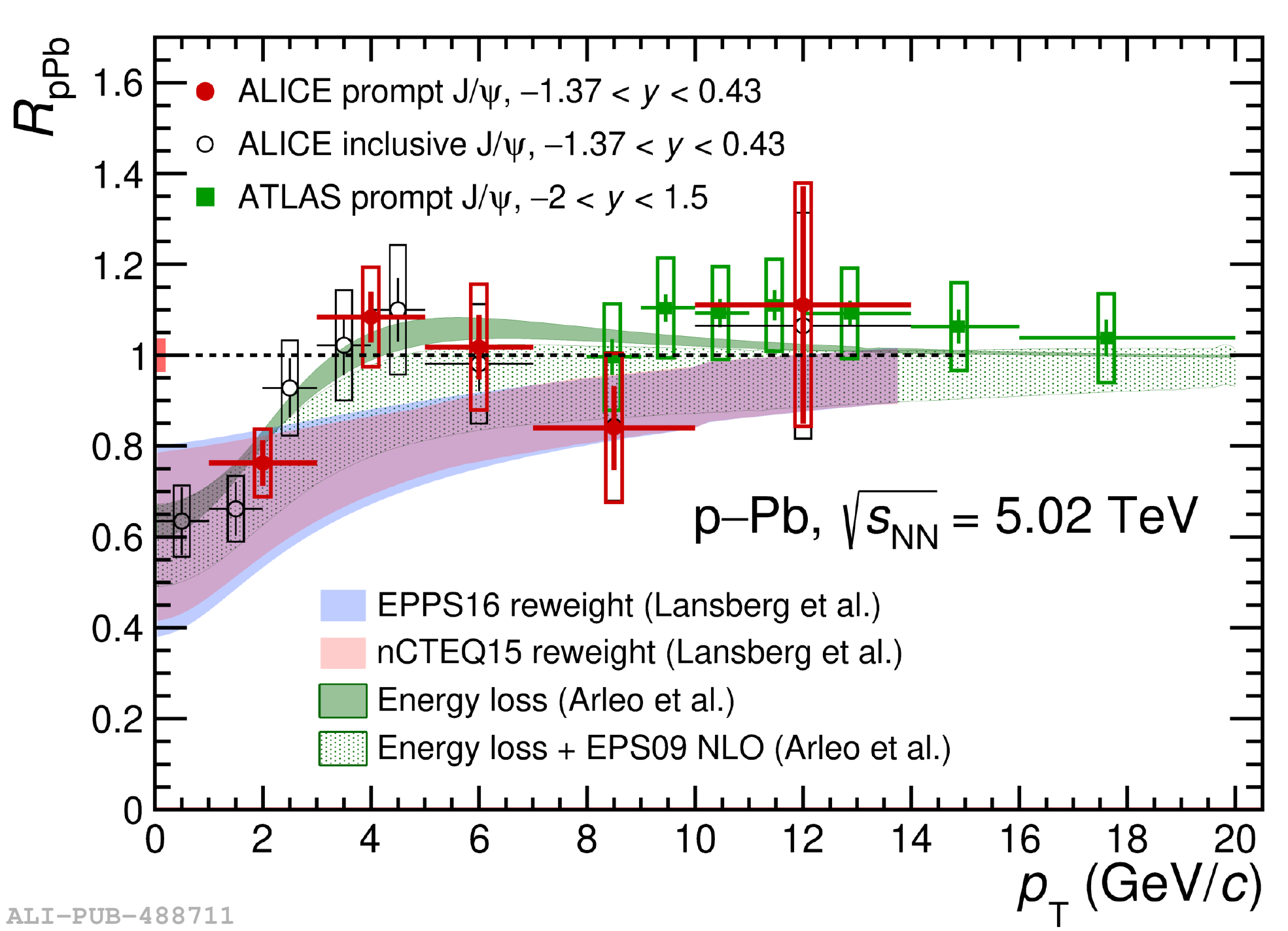}
\includegraphics[width=.45\textwidth]{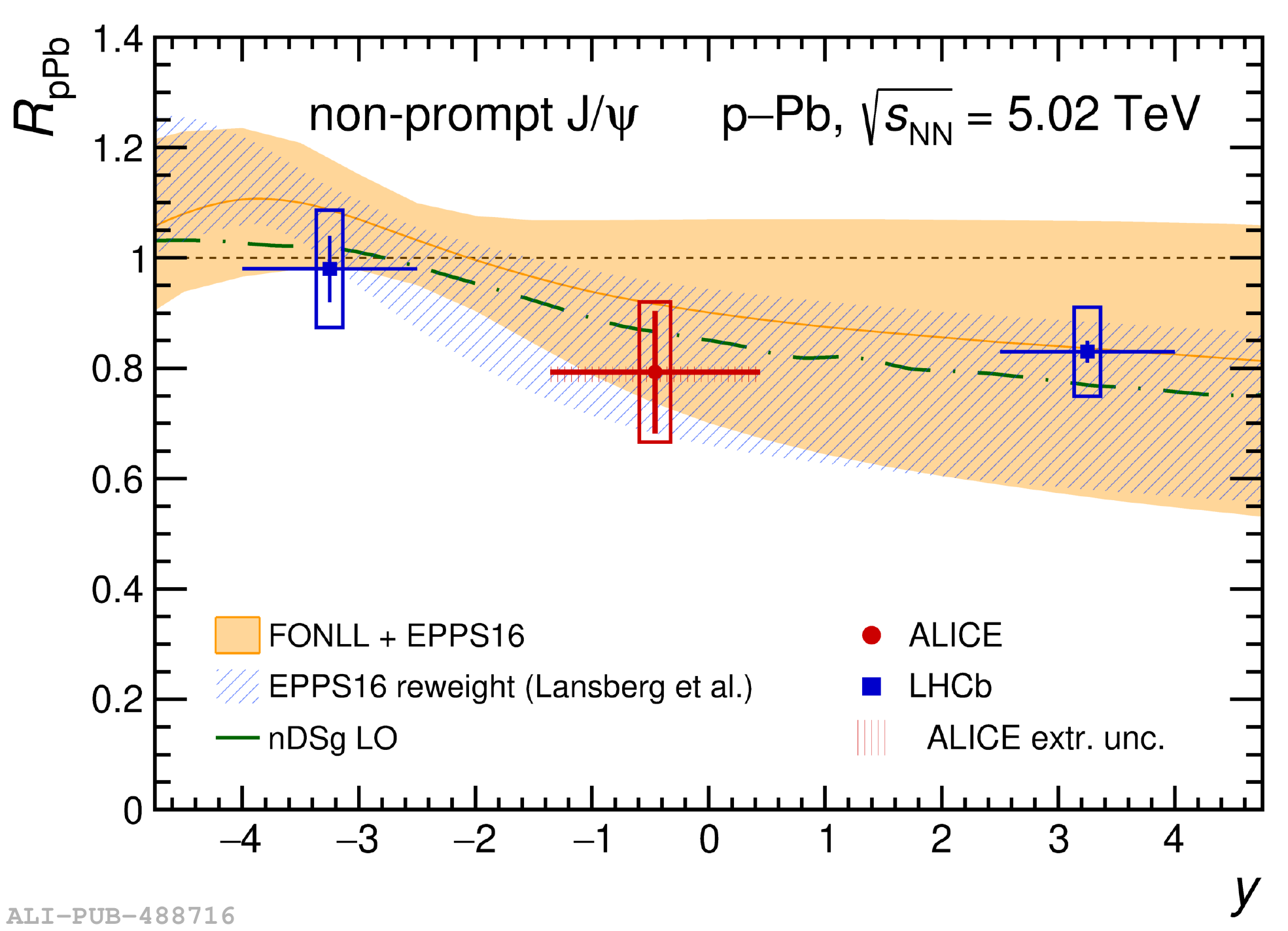}
\caption{Left panel: $R_{\rm pPb}$ of prompt J/$\psi$ as a function of $p_{\rm T}$ along with that of inclusive J/$\psi$ at midrapidity. Right panel: Nuclear modification factor $R_{\rm pPb}$ of non-prompt J/$\psi$ as a function of rapidity~\cite{ALICE:2021lmn}.}
\label{fig-5}       % Give a unique label
\end{figure*}

Recent measurements of inclusive, prompt, and non-prompt J/$\psi$ production in p--Pb collisions at a nucleon--nucleon centre-of-mass energy of $\sqrt{s_{\rm NN}}$ = 5.02 TeV provide additional information on the role of CNM effects~\cite{ALICE:2021lmn}.
The J/$\psi$ mesons are reconstructed in the dielectron decay channel at midrapidity down to a transverse momentum $p_{\rm T}$ = 0. Non-prompt J/$\psi$ mesons, which originate from the decay of beauty hadrons, are separated from promptly produced J/$\psi$ on a statistical basis for $p_{\rm T}>$ 1 GeV/$c$. 
In the left panel of Fig.~\ref{fig-5}, the $R_{\rm pPb}$ of prompt J/$\psi$ is reported  as a function of $p_{\rm T}$ in comparison with ATLAS results~\cite{ATLAS:2017prf}.  Given also the relatively small fraction of J/$\psi$ from b-hadron decays for $p_{\rm T} <$ 14 GeV/$c$, the $R_{\rm pPb}$ of prompt J/$\psi$ is comparable with that of the inclusive J/$\psi$. As shown in the left panel of Fig.~\ref{fig-5}, both trends indicate that the
suppression observed at midrapidity is a low-$p_{\rm T}$ effect, concentrated at  $p_{\rm T}<$ 3 GeV/$c$. The measurements are compared with results from model predictions which embed different CNM effects into prompt J/$\psi$ production~\cite{Kusina:2017gkz}. A calculation including the effects of coherent energy loss~\cite{Arleo:2013zua}, with or without the
introduction of nuclear shadowing effects according to EPS09 nPDF, provides a fairly good description of the measurements.
The measured value of the $p_{\rm T}$-integrated $R_{\rm pPb}$  of non-prompt J/$\psi$ (shown in the right panel of Fig.~\ref{fig-5}) is 0.79 $\pm$ 0.11 (stat.)$\pm$0.13 (syst.)$^{+0.01}_ {-0.02}{\rm (extr.)}$, suggesting the presence of nuclear effects also for the non-prompt component. Within uncertainties, the measurements are compatible with those of the LHCb collaboration~\cite{LHCb:2013gmv} at both forward and backward rapidity. 
All charmonium measurements reported here are in agreement with the mild degree of suppression predicted by models employing nPDFs.

\section{Charm and beauty production in heavy-ion collisions}
\label{sec-3}

\begin{figure*}[ht!]
\centering
\includegraphics[width=.45\textwidth]{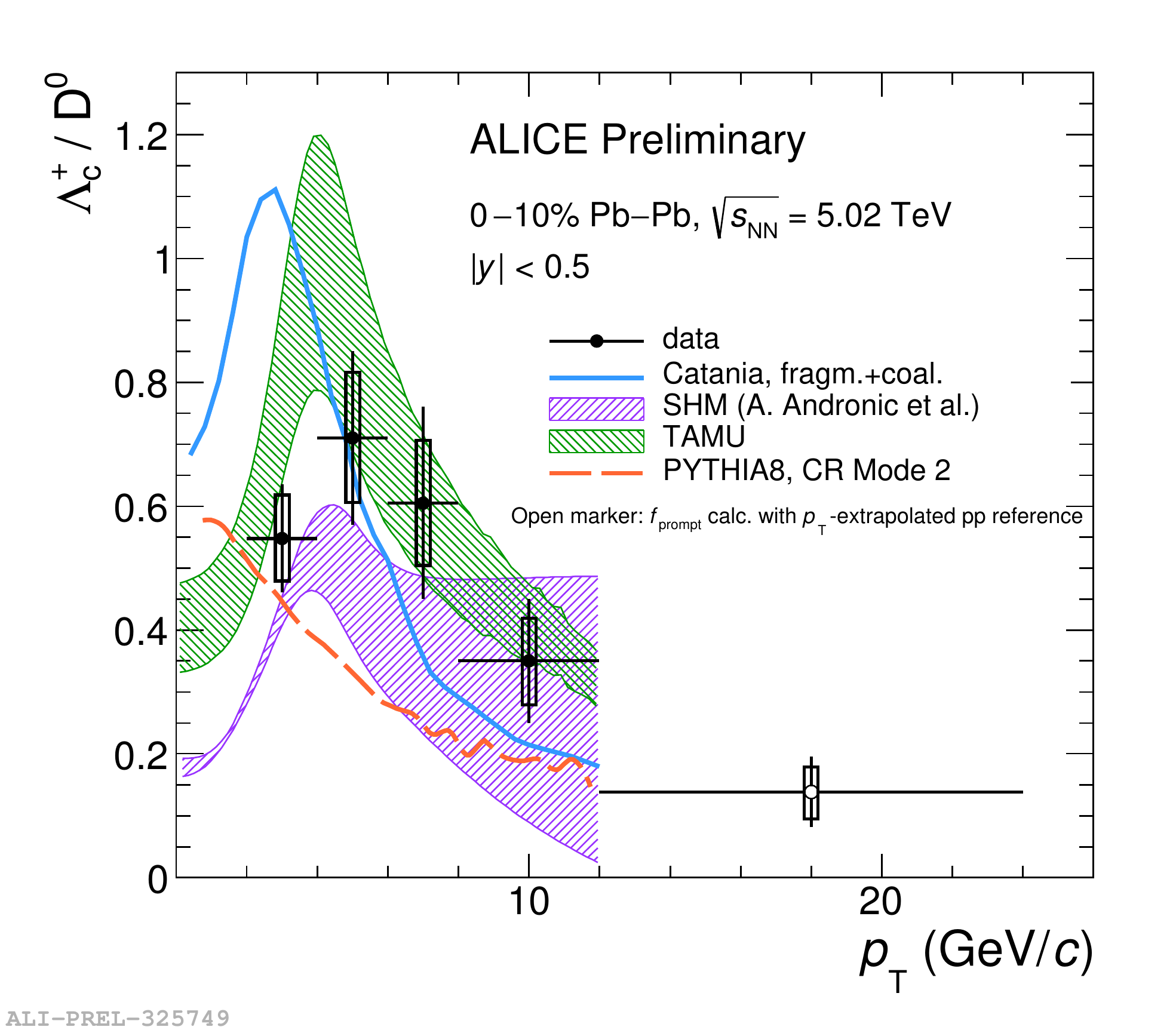}
\includegraphics[width=.4\textwidth]{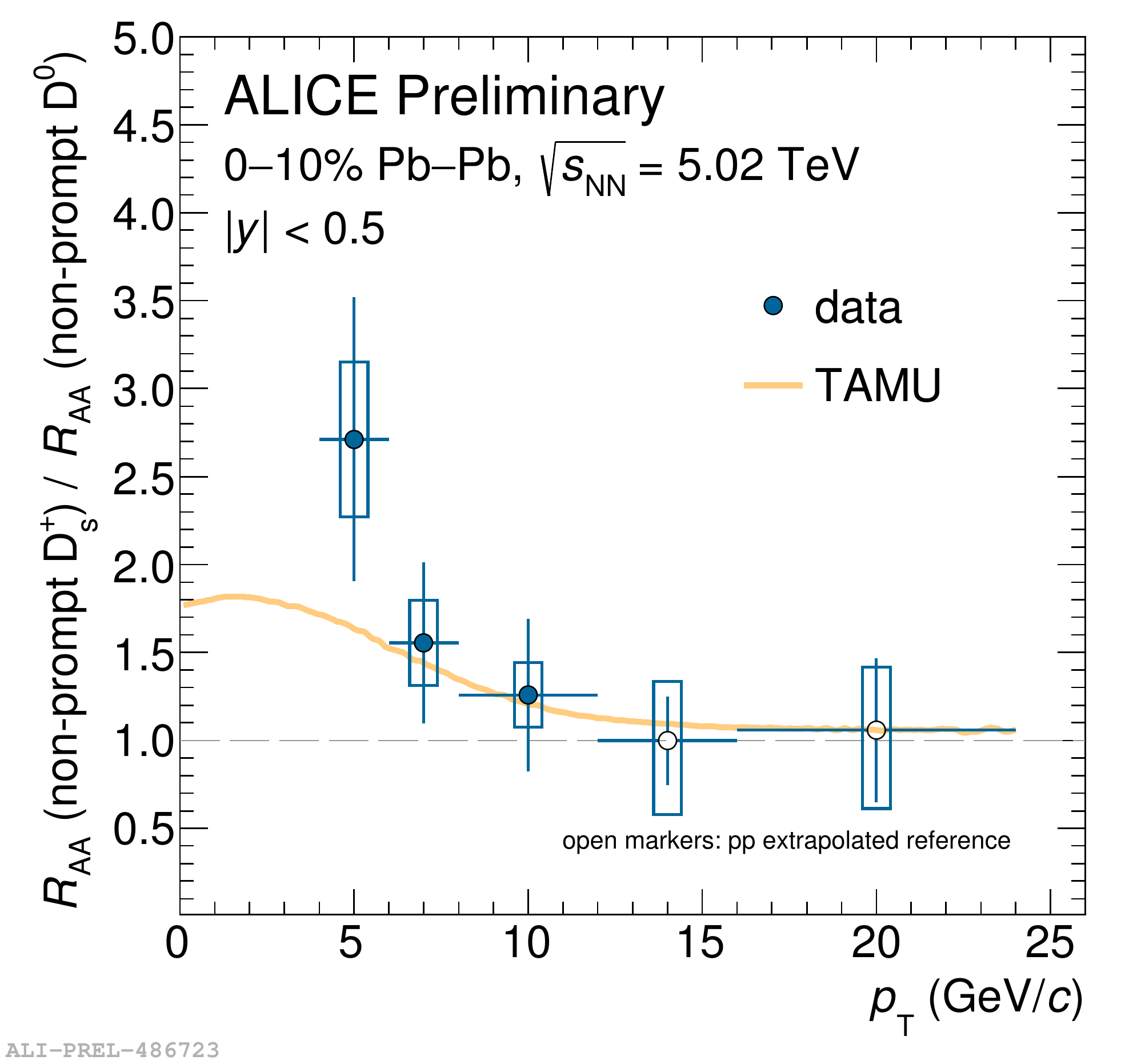}
\caption{Left panel: The $p_{\rm T}$-differential $\Lambda_{\rm c}^+/{\rm D^0}$ yield ratio in pp collisions and in the10\% most central Pb--Pb
collisions at $\sqrt{s_{\rm NN}}$ = 5.02 TeV compared to the predictions of different theoretical calculations. Right panel: Ratio between the $R_{\rm AA}$ of non-prompt ${\rm D_s^+}$ mesons and non-prompt ${\rm D^0}$ mesons in Pb--Pb collisions at $\sqrt{s_{\rm NN}}$= 5.02 TeV for the 0--10\% centrality class compared to the TAMU model~\cite{He:2019vgs}.}
\label{fig-6}       % Give a unique label
\end{figure*}

The $\Lambda_{\rm c}^+/{\rm D^0}$ yield ratio in central Pb--Pb collisions at $\sqrt{s_{\rm NN}}$ = 5.02 TeV extends the study of the hadronisation properties of charm baryons to the extreme conditions realized in central Pb--Pb collisions.  This ratio is larger than the one measured in pp collisions in the interval 4 $<p_{\rm T}< $ 8 GeV/$c$. 
Figure~\ref{fig-6} shows the $p_{\rm T}$-differential $\Lambda_{\rm c}^+/{\rm D^0}$ yield ratio measured in central  collisions compared to three different theoretical predictions: Catania~\cite{Plumari:2017ntm}, TAMU~\cite{He:2019vgs}, and the GSI/Hd+BW statistical hadronisation model~\cite{Andronic:2021erx}.
Theoretical calculations that model the charm-quark transport in the QGP and include hadronisation via both coalescence and fragmentation processes describe the data.

The production of non-prompt ${\rm D^0}$ and ${\rm D_s^+}$ mesons from b-hadron decays was measured at midrapidity ($|y| <$ 0.5) in Pb--Pb collisions at $\sqrt{s_{\rm NN}}$ = 5.02 TeV. The $R_{\rm AA}$ of non-prompt ${\rm D_s^+}$ is measured in the range 2 $<p_{\rm T}<$ 36 GeV/$c$.
A significant suppression up to a factor of about three is observed for $p_{\rm T}>$ 5 GeV/$c$ in the 10\% most central collisions. The data are described by models that include both collisional and radiative processes in the calculation of beauty quark in-medium energy loss and quark coalescence as a hadronisation mechanism.
A higher $R_{\rm AA}$ for non-prompt ${\rm D_s^+}$ mesons than for prompt ${\rm D_s^+}$ mesons is measured at $p_{\rm T} >$ 3 GeV/$c$. This is expected by models in which beauty quarks lose less energy than charm quarks in the QGP medium because of their larger mass. In the right panel of Fig.~\ref{fig-6}, the $R_{\rm AA}$ of non-prompt ${\rm D_s^+}$ mesons was divided by that of non-prompt ${\rm D^0}$ mesons and compared with the same ratio from the TAMU model~\cite{He:2019vgs}. The $R_{\rm AA}$ for non-prompt ${\rm D_s^+}$ mesons is expected to be higher than the one of non-prompt ${\rm D^0}$ mesons due to the enhanced production of ${\rm B_s^0}$ mesons in heavy-ion collisions in case of beauty-quark hadronisation via coalescence. The data are in agreement with the TAMU predictions within the uncertainties.

Further studies on the evolution of heavy-flavour hadronisation across collision systems, including studies as a function of the event multiplicity for rare baryons, will be done in future LHC runs.

\bibliographystyle{woc.bst}   % Remember we use title in the biblio
\bibliography{bibliography}

\begin{thebibliography}{23}

\bibitem{Acharya:2021cqv}
S.~Acharya et~al. (ALICE), JHEP \textbf{05}, 220 (2021), \texttt{2102.13601}

\bibitem{Acharya:2020uqi}
S.~Acharya et~al. (ALICE) (2020), \texttt{2011.06078}

\bibitem{Acharya:2021dsq}
S.~Acharya et~al. (ALICE) (2021), \texttt{2105.05616}

\bibitem{xic13tev}
S.~Acharya et~al. (ALICE) (2021), \texttt{2105.05187}

\bibitem{Cacciari:2012ny}
M.~Cacciari, S.~Frixione, N.~Houdeau, M.L. Mangano, P.~Nason, G.~Ridolfi, JHEP
  \textbf{10}, 137 (2012), \texttt{1205.6344}

\bibitem{Kramer:2017gct}
G.~Kramer, H.~Spiesberger, Nucl. Phys. B \textbf{925}, 415 (2017),
  \texttt{1703.04754}

\bibitem{Helenius:2018uul}
I.~Helenius, H.~Paukkunen, JHEP \textbf{05}, 196 (2018), \texttt{1804.03557}

\bibitem{Christiansen:2015yqa}
J.R. Christiansen, P.Z. Skands, JHEP \textbf{08}, 003 (2015),
  \texttt{1505.01681}

\bibitem{Song:2018tpv}
J.~Song, H.h. Li, F.l. Shao, Eur. Phys. J. C \textbf{78}, 344 (2018),
  \texttt{1801.09402}

\bibitem{Minissale:2020bif}
V.~Minissale, S.~Plumari, V.~Greco (2020), \texttt{2012.12001}

\bibitem{He:2019tik}
M.~He, R.~Rapp, Phys. Lett. B \textbf{795}, 117 (2019), \texttt{1902.08889}

\bibitem{Skands:2014pea}
P.~Skands, S.~Carrazza, J.~Rojo, Eur. Phys. J. C \textbf{74}, 3024 (2014),
  \texttt{1404.5630}

\bibitem{ALICE:2021dhb}
S.~Acharya et~al. (ALICE) (2021), \texttt{2105.06335}

\bibitem{dEnterria:2016yhy}
D.~d'Enterria, A.M. Snigirev, Eur. Phys. J. C \textbf{78}, 359 (2018),
  \texttt{1612.08112}

\bibitem{Czakon:2013goa}
M.~Czakon, P.~Fiedler, A.~Mitov, Phys. Rev. Lett. \textbf{110}, 252004 (2013),
  \texttt{1303.6254}

\bibitem{ALICE:2021lmn}
S.~Acharya et~al. (ALICE) (2021), \texttt{2105.04957}

\bibitem{ATLAS:2017prf}
M.~Aaboud et~al. (ATLAS), Eur. Phys. J. C \textbf{78}, 171 (2018),
  \texttt{1709.03089}

\bibitem{Kusina:2017gkz}
A.~Kusina, J.P. Lansberg, I.~Schienbein, H.S. Shao, Phys. Rev. Lett.
  \textbf{121}, 052004 (2018), \texttt{1712.07024}

\bibitem{Arleo:2013zua}
F.~Arleo, R.~Kolevatov, S.~Peign\'e, M.~Rustamova, JHEP \textbf{05}, 155
  (2013), \texttt{1304.0901}

\bibitem{LHCb:2013gmv}
R.~Aaij et~al. (LHCb), JHEP \textbf{02}, 072 (2014), \texttt{1308.6729}

\bibitem{Plumari:2017ntm}
S.~Plumari, V.~Minissale, S.K. Das, G.~Coci, V.~Greco, Eur. Phys. J. C
  \textbf{78}, 348 (2018), \texttt{1712.00730}

\bibitem{He:2019vgs}
M.~He, R.~Rapp, Phys. Rev. Lett. \textbf{124}, 042301 (2020),
  \texttt{1905.09216}

\bibitem{Andronic:2021erx}
A.~Andronic, P.~Braun-Munzinger, M.K. K\"ohler, A.~Mazeliauskas, K.~Redlich,
  J.~Stachel, V.~Vislavicius, JHEP \textbf{07}, 035 (2021), \texttt{2104.12754}

\end{thebibliography}

%
% BibTeX or Biber users please use (the style is already called in the class, ensure that the "woc.bst" style is in your local directory)
% \bibliography{name or your bibliography database}
%
% Non-BibTeX users please use
%
%\begin{thebibliography}{}
%
% and use \bibitem to create references.
%
%\bibitem{RefJ}
% Format for Journal Reference
%Journal Author, Journal \textbf{Volume}, page numbers (year)
% Format for books
%\bibitem{RefB}
%Book Author, \textit{Book title} (Publisher, place, year) page numbers
% etc
%\end{thebibliography}

\end{document}